\documentclass[referee]{aa} 
\usepackage{graphicx}
\begin{document}
   \title{A New Three-Parameter Correlation for Gamma-ray Bursts with a Plateau Phase in the Afterglow}


   \author{M. Xu$^{1, 2}$
          \and
          Y. F. Huang$^{1, 3}$
          }

   \institute{$^1$ Department of Astronomy, Nanjing University, Nanjing 210093, China\\
              $^2$ Department of Physics, Yunnan University, Kunming 650091, China \\
              $^3$ Key Laboratory of Modern Astronomy and Astrophysics (Nanjing University),
              Ministry of Education, China \\
              \email{hyf@nju.edu.cn}
             }

   \date{Received 00 00, 0000; accepted 00 00, 0000}


  \abstract
    {}
   {Gamma ray bursts (GRBs) have great advantages for their huge burst energies, luminosities
and high redshifts in probing the Universe. A few interesting luminosity
correlations of GRBs have been used to test cosmology models.
Especially, for a subsample of long GRBs with known redshifts and a plateau phase
in the afterglow, a correlation between the end time of the plateau phase (in the GRB rest
frame) and the corresponding X-ray luminosity has been found.
   }
  {In this paper, we re-analyze the subsample and found that a significantly tighter
correlation exists when we add a third parameter, i.e. the isotropic $\gamma$-ray energy
release, into the consideration. We use the Markov chain Monte Carlo techniques to get
the best-fit coefficients.}
   {A new three-parameter correlation is found for the GRBs with an obvious
plateau phase in the afterglow. The best fit correlation is found to be
$L_{\rm X} \propto T_{\rm a}^{-0.87} E_{\gamma,\rm iso}^{0.88}$.
Additionally, both long and intermediate duration GRBs
are consistent with the same three-parameter correlation equation.
   }
   {It is argued that
the new three-parameter correlation is consistent with the hypothesis that the subsample
of GRBs with a plateau phase in the afterglow be associated with the birth of
rapidly rotating magnetars, and that the plateau be due to the continuous energy-injection
from the magnetar. It is suggested that the newly born millisecond magnetars associated with
GRBs might provide a good standard candle in the Universe.}


\authorrunning{M. Xu \& Y. F. Huang}

\titlerunning{L-T-E Correlation for GRBs with Plateau Phase}

   \maketitle
%

\section{Introduction}

Gamma-ray busts (GRBs) are one of the most powerful and energetic explosive events
in the Universe. The observations of GRBs up to redshifts higher than 8
(Salvaterra et al. 2009; Cucchiara et al. 2011) make
GRBs to be among the farthest known astrophysical sources.
Taking their considerable event rate into consideration, GRBs may be good candidates
that can be used to probe our Universe. Several interesting correlations
have been suggested for GRBs (Amati et al. 2002; Norris et al. 2000;
Ghirlanda et al. 2004a; Liang \& Zhang 2005; Dainotti et al. 2010;
Qi \& Lu 2010). Based on them, the cosmology parameters have been tentatively
constrained (e.g., Fenimore \& Ramirez-Ruiz 2000; Schaefer 2003, 2007;
Dai et al. 2004; Ghirlanda et al. 2004b, 2006; Amati et al. 2008;
Wang \& Dai 2006; Dainotti et al. 2008; Wang et al. 2009, 2011).

To derive a meritorious constraint on the cosmology parameters, the most important thing
is to find a credible standard candle relation for GRBs. Currently, no such a relation
can be established when all GRBs are involved (Butler et al. 2009; Yu et al. 2009).
The reason may be that different GRBs should be produced via various mechanisms.
Interestingly, for a subsample of long GRBs with known redshifts and with a plateau
phase in the afterglow, an anti-correlation has been reported to exist between
the end time of the plateau phase ($T_{\rm a}$, measured in the GRB rest frame) and
the corresponding X-ray luminosity ($L_{\rm X}$) at that moment (Dainotti et al. 2010,
hereafter D2010). In this paper, we call Dainotti et al.'s two parameter correlation as
the L-T correlation. The intrinsic scatter of this correlation is still too large to
be directly applied as a redshift estimator (Dainotti et al. 2011). Additionally,
normal long duration GRBs and the intermediate duration GRBs do not obey the
same correlation equation (D2010), and the intermediate class seem to be
more scattered in the plot.

In this study, we have tried to add a third parameter, i.e. the isotropic $\gamma$-ray energy
release ($E_{\gamma, \rm iso}$), into the correlation. We find that the new three-parameter
correlation (designated as the L-T-E correlation) is much tighter than the previous
L-T correlation. It is also obeyed by both the long GRBs and the intermediate calss.
The L-T-E correlation may hopefully give a better measure for our Universe.
In Section 2, we describe our GRB sample and the method of data analysis.
Our results are presented in Section 3. Section 4 is our discussion and conclusions.

\section{Sample \& Data analysis}

According to $Swift$ observations, many GRBs show a plateau phase in the early
afterglow, prior to the normal power-law decay phase (Zhang et al. 2006;
Nousek et al. 2006). In this study, we will mainly concentrate on the GRBs
with such a characteristics.
All our GRBs are taken from Dainotti et al.'s sample (D2010). In D2010's data
table, totally 77 GRBs are initially included, with known redshift and with a
plateau phase in the afterglow light curve. After removing the intermediate class
GRBs and some GRBs with relatively large errors, they finally limited their major
statics to only 62 long GRBs. Here, we have re-selected the events by taking into
account the following three criterions in our studies: (1) the plateau should be obvious
(GRBs 050318, 050603, 060124, 060418, 061007, 070518 and 071031 are removed
by us, since their phateau phase is not clear enough.);
(2) the data in the plateau phase should be rich enough to show the profile of the
plateau and the end time of the plateau as well (GRBs 050820A, 060512, 060904 and
060124 are removed by us due to this constraint.);
and (3) there should be no flares during the plateau phase, since flares may affect the
shape of the plateau light curve and lead to errors in the quantities that we are
interested in (GRBs 050904, 050908, 060223A and 060526 are removed by us according
to this condition.). As a result, our ``golden sample'' is consisted of
55 events in total, i.e., 47 long GRBs and 8 intermediate class GRBs
(Intermediate class GRB is characterized by a short initial burst followed by an extended low intensity
emission phase; Norris et al. 2006).
The redshifts of our sample range from 0.08 to 8.26.

For the end times of the plateau phase ($T_{\rm a}$, in the GRB rest frame) and the
X-ray afterglow luminosities at that moment ($L_{\rm X} \equiv L_{\rm X}(T_{\rm a})$),
we use the values of D2010. In D2010, $T_{\rm a}$ is derived through a phenomenological
fitting model (Willingale et al. 2007), and $L_{\rm X}$ is derived from the following equation,
\begin{equation}
L_{\rm X}=\frac{4 \pi D_{\rm L}^{2}(z)F_{\rm X}}{(1+z)^{1-\beta_{\rm a}}},
\end{equation}
where $z$ is the redshift, $D_{\rm L}(z)$ is the luminosity distance, $F_{\rm X}$ is the
observed flux by $Swift-XRT$ at the end time of the plateau phase, and $\beta_{\rm a}$ is the spectral
index of the X-ray afterglow (Evans et al. 2009).

The isotropic $\gamma$-ray energy release in the prompt emission phase is
\begin{equation}
E_{\gamma,\rm iso}=4\pi D_{\rm L}^{2}(z)S_{\rm bolo}/(1+z),
\end{equation}
where $S_{\rm bolo}$ is the bolometric fluence, and can be taken from Wang et al. (2011).
In the study of Wang et al. (2011), $S_{\rm bolo}$ is calculated from the observed energy
spectrum $\Phi(E)$ as (Schaefer 2007):
\begin{equation}
S_{\rm bolo}=S\times\frac{\int_{1/(1+z)}^{10^{4}/(1+z)}E\Phi(E)dE}
 {\int_{E_{\rm min}}^{E_{\rm max}}E\Phi(E)dE},
\end{equation}
where $S$ is the observed fluence in units of $\rm erg\cdot cm^{-2}$ for each GRB,
and ($E_{\rm min},~E_{\rm max}$) are the detector threshold.
The energy spectrum $\Phi(E)$ is assumed to be the Band function (Band et al. 1993),
\begin{equation}
\Phi(E)= \left \{
 \begin{array}{ll}
  AE^{\alpha}e^{-(2+\alpha)E/E_{\rm peak}}~~E\leq[(\alpha-\beta)/(2+\alpha)]E_{\rm peak} \\
  BE^{\beta}~~~~~~~~~~~~~~~~~~~~~~~~~~~~~\rm otherwise
 \end{array}
\right.
\end{equation}
where $E_{\rm peak}$ is the peak energy of the spectrum,
and $\alpha$, $\beta$ are the power-law indices for photon energies below or above the
break energy respectively. At last, the complete data set of all our 55 GRBs are
shown in Table 1, where the error bars are $1 \sigma$ range.

We investigate if an intrinsic correlation exists between the three
parameters of $L_{\rm X},~T_{\rm a}$ and $E_{\gamma,\rm iso}$ as following,
\begin{equation}
{\rm log}(\frac{L_{\rm X}}{10^{47}\rm erg\cdot s^{-1}})=\it a+b{\rm ~log}(\frac{T_{\rm a}}{\rm 10^{3}s})
  +\it c{\rm ~log}(\frac{E_{\gamma,\rm iso}}{\rm 10^{53}erg}),
\end{equation}
where $a,~b$, and $c$ are constants to be determined from the fit to
the observational data. In this equation, $a$ is the constant of the
intercept. $b$ and $c$ are actually the power-law indices of time
and energy when we approximate $L_{\rm X}$ as power-law functions of
$T_{\rm a}$ and $E_{\gamma,\rm iso}$. Due to the complexity of GRB sampling,
an intrinsic scattering parameter, $\sigma_{\rm int}$, is introduced in our
analysis, as is usually done by other researchers
(Reichart 2001; Guidorzi et al. 2006; Amati et al 2008).
This extra variable that follows a normal distribution of $N(0,~\sigma_{\rm int}^{2})$
is engaged to represent all the contribution to $L_{\rm X}$ from other
unknown hidden variables.

To derive the best fit to the observational data with the above three-parameter
correlation, we use the method presented in D$'$Agostini (2005).
Here, for simplify, we first define $x_{1}={\rm log}(\frac{T_{\rm a}}{10^{3}\rm s})$,
$x_{2}={\rm log}(\frac{E_{\gamma,\rm iso}}{10^{53}\rm erg})$,
and $y={\rm log}(\frac{L_{\rm X}}{10^{47}\rm erg/s})$.
The joint likelihood function for the coefficients of $a,~b,~c$ and $\sigma_{\rm int}$ is
(D'Agostini 2005)
\begin{equation}
\begin{array}{rcl}
\mathcal{L}(a,b,c,\sigma_{\rm int}) \propto \displaystyle{\prod_{i}}
  \frac{1}{\sqrt{\sigma_{\rm int}^{2}+\sigma_{y_{i}}^{2}+b^{2}\sigma_{x_{1,i}}^{2}+c^{2}\sigma_{x_{2,i}}^{2}}}\\
  \times \exp [-\frac{(y_{i}-a-bx_{1,i}-cx_{2,i})^{2}}
    {2(\sigma_{\rm int}^{2}+\sigma_{y_{i}}^{2}+b^{2}\sigma_{x_{1,i}}^{2}+c^{2}\sigma_{x_{2,i}}^{2})}],
\end{array}
\end{equation}
where $i$ is the corresponding serial number of GRBs in our sample.

In order to get the best-fit coefficients, the so called Markov chain
Monte Carlo techniques are used in our calculations. For each Markov
chain, we generate $10^{6}$ samples according to the likelihood
function. Then we derive the the coefficients of $a,~b,~c$ and $\sigma_{\rm int}$ according to
the statistical results of the samples.

Our likelihood function can also be conveniently applied to the
two-parameter L-T correlation case studied by D2010, by simply
taking $c=0$. We have checked our method by comparing our result
for the L-T correlation with that of D2010. The results are generally
consistent, which proves the reliability of our codes.

\section{Results}

In our study, we assume a flat $\Lambda \rm CDM$ cosmology
with $H=69.7~\rm km\cdot s^{-1}\cdot Mpc^{-1}$
and $\Omega_{\rm M}=0.291$ (the same values as D2010).
By using the method described in Section 2, we find
that the best-fit correlation between $L_{\rm X}$, $T_{\rm a}$ and
$E_{\gamma,\rm iso}$ is
\begin{equation}
\begin{array}{rcl}
{\rm log}(\frac{L_{\rm X}}{10^{47}{\rm erg/s}})=1.17-0.87~{\rm log}(\frac{T_{\rm a}}{10^{3}\rm s})
  +0.88{\rm ~log}(\frac{E_{\gamma,\rm iso}}{10^{53}\rm erg}).
\end{array}
\end{equation}
Figure 1 shows the above correlation.
In this figure, the solid line is plotted from Eq. (7), and
the points represent the 55 GRBs of our sample (the filled points correspond
to the 47 long GRBs and the hollow square points correspond to the 8 intermediate class GRBs).
It is clearly shown that this three-parameter correlation is tight for all the
55 GRBs.

\begin{figure}
\label{fig1}
 \includegraphics[width=0.8\textwidth]{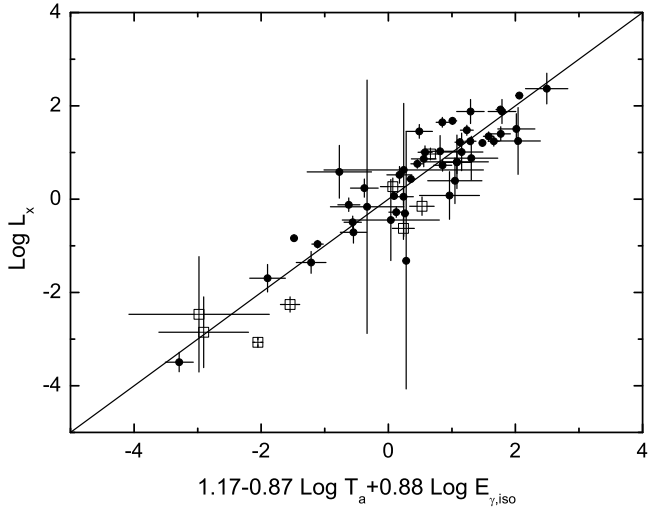}
 \caption{The best-fit correlation between $L_{\rm X}$, $T_{\rm a}$ and
$E_{\gamma,\rm iso}$ for our ``golden sample''.
Y-axis is the X-ray luminosity at the end time of the
plateau phase, i.e. $L_{\rm X}$, in units of $10^{47}$ erg/s. Note that
the X-axis is a combined quantity of $T_{\rm a}$ (in units of $10^3$ s)
and $E_{\gamma,\rm iso}$ (in units of $10^{53}$ erg), i.e.
$1.17-0.87 \log {T_{\rm a}} +0.88 \log E_{\gamma,\rm iso}$. The filled points
correspond to the observed data of 47 long GRBs and the hollow square points correspond to
the 8 intermediate class GRBs. The solid line is plotted from Eq. (7), which
is the best fit of the 55 observational data points.}
\end{figure}

Comparing Eqs. (5) and (7), we find that the best values for the
constants of $a$, $b$, and $c$ in Eq.~(5) are $a = 1.17$,
$b = -0.87$, and $c = 0.88$ respectively. Figure 1 also clearly
shows that there is still obvious scatter in the L-T-E correlation.
To give a quantitative description of the scatter, we need to derive
the $1 \sigma$ errors of these constants.

The probability distributions of these constants as well as the intrinsic
scattering parameter ($\sigma_{\rm int}$) are displayed in Figure 2.
From this figure, we find that the probability distributions of these
coefficients can be well fitted by Gauss functions. So we can easily get
the $1\sigma$ error bars for these parameters.
Actually, the best values and the $1\sigma$
errors for the coefficients are $a = 1.17 \pm 0.09$, $b = -0.87 \pm 0.09$,
$c = 0.88 \pm 0.08$, and $\sigma_{\rm int} = 0.43 \pm 0.05$, respectively.

\begin{figure}
 \begin{minipage}[b]{0.5\textwidth}
   \includegraphics[width=0.7\textwidth]{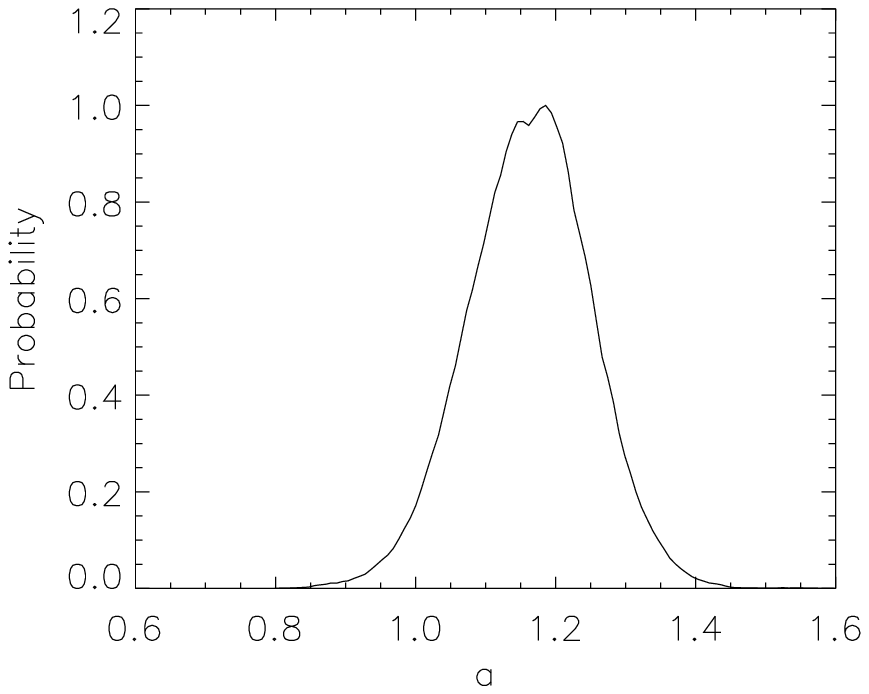}
   \includegraphics[width=0.7\textwidth]{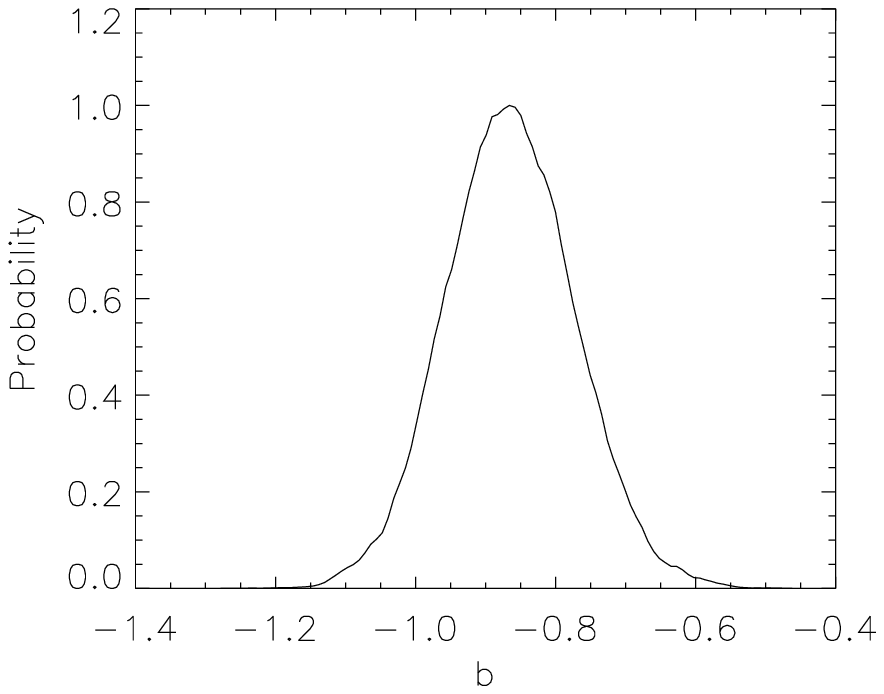}
 \end{minipage}
\phantom{0}
 \begin{minipage}[b]{0.5\textwidth}
   \includegraphics[width=0.7\textwidth]{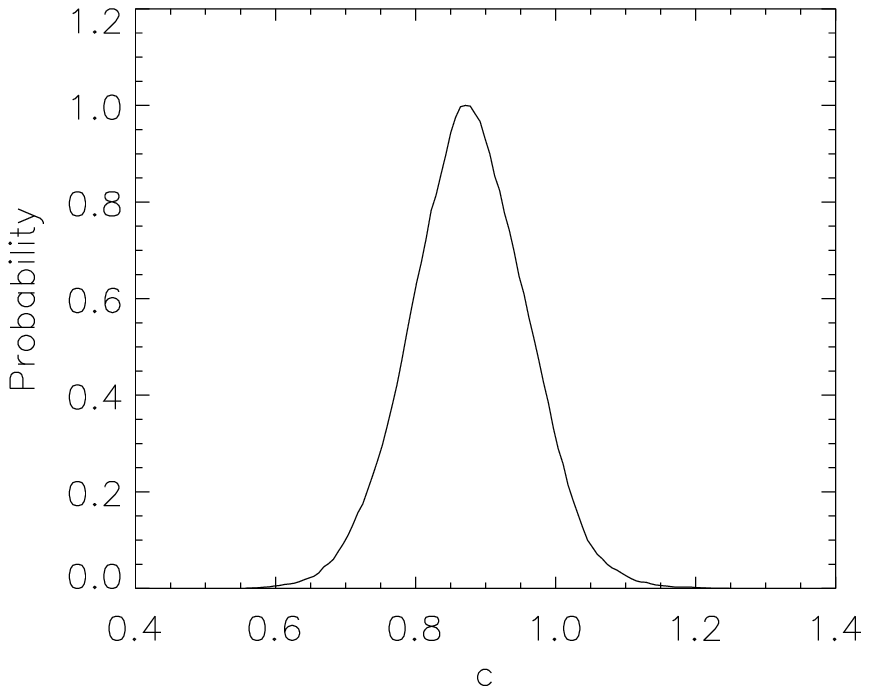}
   \includegraphics[width=0.7\textwidth]{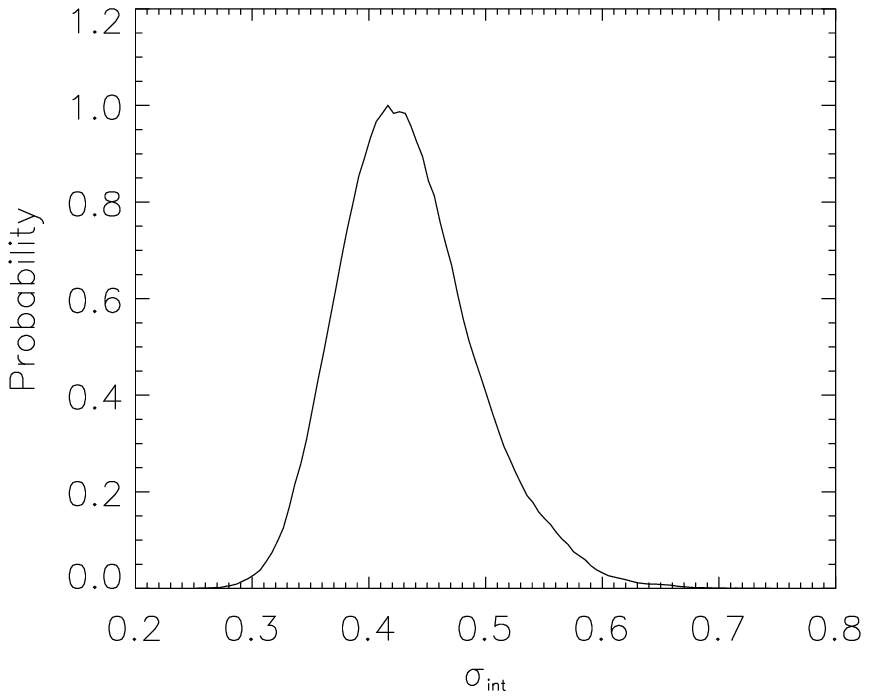}
 \end{minipage}
\caption{The probability distributions of the constants of $a$ (upper left panel),
$b$ (upper right panel), $c$ (lower left panel) in Eq. (5), and the
probability distribution of the intrinsic scattering parameter $\sigma_{\rm int}$
(lower right panel). According to these panels, the best values and the $1\sigma$
errors for the coefficients are $a = 1.17 \pm 0.09$, $b = -0.87 \pm 0.09$,
$c = 0.88 \pm 0.08$, $\sigma_{\rm int} = 0.43 \pm 0.05$, respectively.
\label{fig2}}
\end{figure}

We have also explored the three-parameter correlation for
all the 77 GRB events listed in D2010, using the same analytical method as
for our ``golden sample'' of 55 GRBs.
The best fit result is shown in Figure 3. The best parameter values and the $1 \sigma$
errors for the coefficients are $a = 0.81 \pm 0.07$, $b = -0.91 \pm 0.09$,
$c = 0.59 \pm 0.05$, and $\sigma_{\rm int} = 1.15 \pm 0.12$.
Comparing with the result of the ``golden sample'', although there is still an
obvious correlation among $L_{\rm X}$, $T_{\rm a}$ and $E_{\gamma,\rm iso}$ for all the 77 GRBs,
the intrinsic scatter of the L-T-E correlation is much larger now.
However, it is very important to note that
we exclude the 22 samples because they are most likely
not physically belonging to the same group as the ``golden sample''
(for example, many of them do not have an obvious plateau phase), as judged from
the three criterions in Section 2.

\begin{figure}
\label{fig2}
 \includegraphics[width=0.8\textwidth]{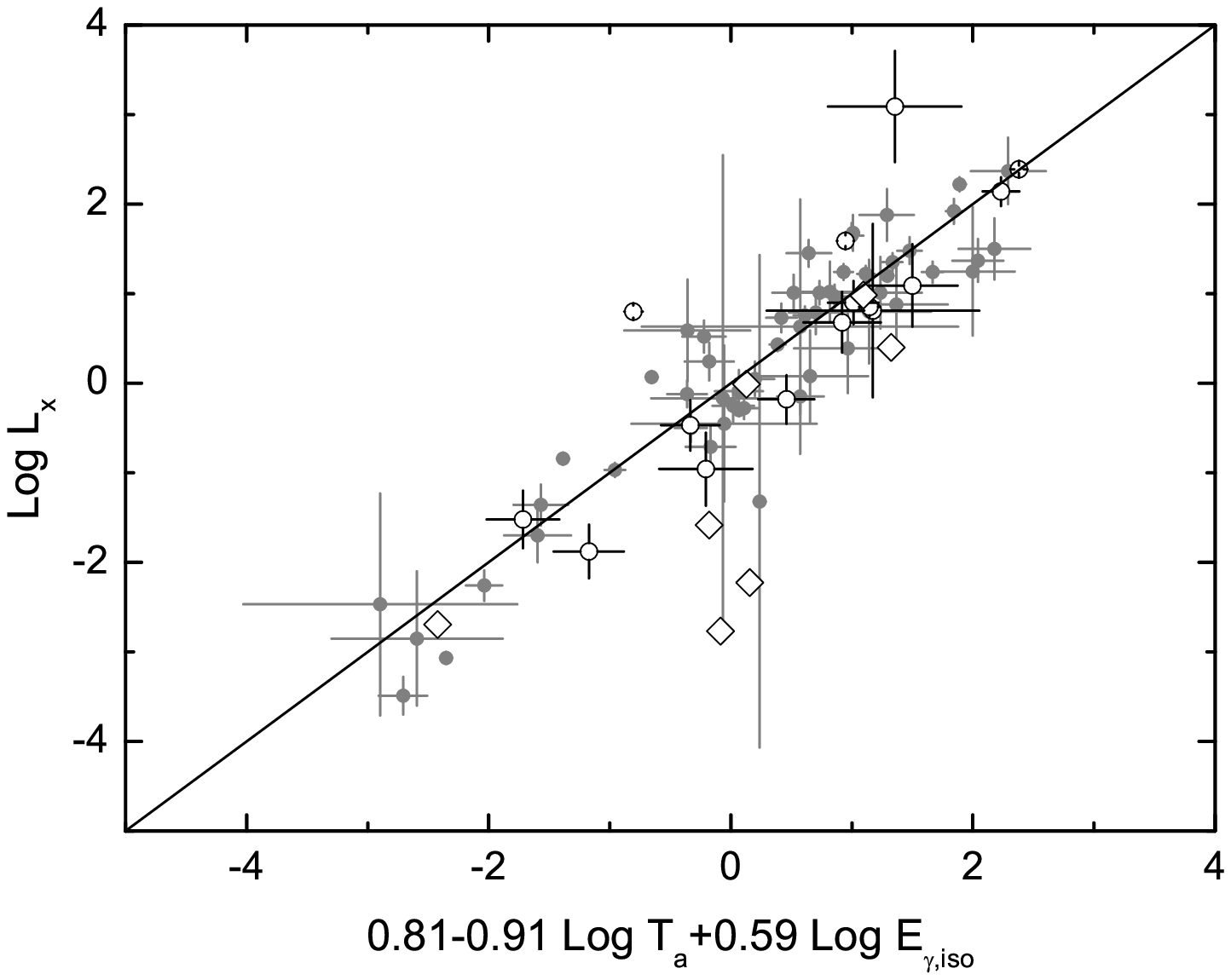}
 \caption{The best-fit correlation between $L_{\rm X}$, $T_{\rm a}$ and
$E_{\gamma,\rm iso}$ for all the 77 GRBs of D2010. The units of all physical quantity
are the same as Figure 2. The X-axis is a combined quantity of
$0.81-0.91 \log {T_{\rm a}} +0.59 \log E_{\gamma,\rm iso}$. The filled points
correspond to the observed data of 55 ``golden'' GRBs with error bars. The hollow diamonds
correspond to 7 GRBs with too large error bars to be plotted in the figure, and the hollow
circles correspond to other 15 discarded events.
The solid line is the best fit for all the 77 data points.}
\end{figure}

In order to directly compare with the L-T correlation suggested by D2010, we have also fit
the two-parameter correlation for our sample. The best-fit equation is
\begin{equation}
\begin{array}{rcl}
{\rm log}(\frac{L_{\rm X}}{10^{47}\rm erg/s})=(0.78\pm0.14)
  -(1.16\pm0.16){\rm ~log}(\frac{T_{\rm a}}{10^{3}\rm s}).
\end{array}
\end{equation}
This equation is consistent with the L-T correlation derived in D2010.
Comparing Eq. (8) with Eq. (7) and from Figure 2, we find that the error
bars of the constants in Eq. (8) (i.e. the L-T correlation)
are generally significantly larger than those of Eq. (7) (i.e. the L-T-E
correlation). Additionally, in the two-parameter fitting of Eq. (8), the
intrinsic scatter is $0.85 \pm 0.10$, which is also markedly larger than
that in the three-parameter correlation case ($0.43 \pm 0.05$). From the
comparison, we see that the L-T-E correlation is really significantly tighter
than the L-T correlation.

For our GRB sample, we additionally find that the correlation coefficient of
our L-T-E statistics is $r=0.92$ and the chance probability is
$P=1.05 \times 10^{-20}$. On the contrary, the correlation
coefficient of the L-T statistics of the same sample is $r=-0.73$
and the corresponding chance probability is $P = 5.55 \times 10^{-8}$.
This also shows that the L-T-E correlation is much tighter than the
L-T correlation.

\section{Discussion and Conclusions}

In this paper, a new three-parameter correlation is found for the GRBs with an obvious
plateau phase in the afterglow. This L-T-E correlation is tighter
than the L-T correlation reported in D2010. It has been shown that the intrinsic scattering
of our L-T-E correlation is significantly smaller than that of the L-T correlation, and
the correlation coefficient is correspondingly larger. However, we note that the
intrinsic scatter of the L-T-E correlation is still larger than that of some correlations
derived from prompt GRB emission (Guidorzi et al. 2006; Amati et al. 2008). In the future,
more samples and more delicate selections might help to improve the result.

The plateau phase (or the shallow decay segment) is an interesting characteristics
of many GRB afterglows (Zhang et al. 2006; Nousek et al. 2006). This phenomenon can be
explained as continuous energy injection from the central engine after the prompt burst
(Rees \& M\'{e}sz\'{a}ros 1998; Dai \& Lu 1998; Zhang \& M\'{e}sz\'{a}ros 2001;
Dai 2004; Kobayashi \& Zhang 2007; Yu \& Dai 2007; Xu et al. 2009;
Yu et al. 2010; Dall$'$Osso et al. 2011),
or by the two component models (Corsi \& M{\'e}sz{\'a}ros 2009),
or by structured jets (Eichler \& Granot 2006; Granot et al. 2006; Panaitescu 2007; Yamazaki 2009; Xu \& Huang 2010),
or even as due to dust scattering (Shao \& Dai 2007; Shao et al. 2008).
According to our L-T-E correlation (Eq. (7)), the X-ray luminosity at the end time
of the plateau can be expressed as a function of the end time and the isotropic $\gamma$-ray
energy release as,
\begin{equation}
L_{\rm X} \propto T_{\rm a}^{-0.87 \pm 0.09} E_{\gamma,\rm iso}^{0.88 \pm 0.08}.
\end{equation}
We believe that this relation can give useful constraint on the underlying physics.

For the energy injection model, a natural mechanism is the dipole radiation from
the spinning down of a magnetar at the center of the fireball. Note
that the injected energy may not be Poynting flux, but can be electron-positron pairs
(Dai 2004). These pairs interact with the fireball material,
leading to the formation of a relativistic wind bubble.
When the energy injection dominates the dynamical evolution of the external
shock, the afterglow intensity should naturally be proportional to the
energy injection power. So, $L_{\rm X}$ is actually a measure of the energy
injection rate. According to Eq. (9), $L_{\rm X}$ is roughly inversely proportional
to the timescale of the energy injection, $T_{\rm a}$. It hints that the energy
reservoir should be roughly a constant. This is consistent with the
energy injection model, which usually assumes that the central engine is
a rapidly rotating millisecond magnetar. In different GRBs, the surface
magnetic field intensities of the central magnetars may be quite different,
leading to various energy injection luminosities and energy injection timescales.
But the total energy available for energy injection is relatively constant
(about rotational energy of the magnetar).
It is mainly constrained by the limiting angular velocity of the magnetar,
which again is determined by the equation of state of neutron stars.
Additionally, according to Dai (2004), in order to produce an obvious plateau in
the afterglow lightcurve, the total injected energy must be comparable to the original
fireball energy (which may be comparable to $E_{\gamma,\rm iso}$). This
requirement is again roughly consistent with the item of
$E_{\gamma, \rm iso}^{0.88 \pm 0.08}$ in Eq. (9).
Based on the above analysises, we argued that the L-T-E correlation strongly
supports the energy injection model of magnetars.
It also indicates that the newly born millisecond magnetars associated with
GRBs provide a good standard candle in our Universe. Thus the
L-T-E correlation may potentially be used to test the cosmological models.

Our sample contains 47 long GRBs and 8 intermediate class GRBs. From Figure 1, we see that both
of these two classes are consistent with the same L-T-E correlation. Howerer, note that they behave very
differently in frame work of the two-parameter L-T correlation. This is another important
advantage of our three-parameter correlation. It indicates that magnetars may also form in
intermediate class GRBs, and their limiting spinning is just similar to those magnetars
born in long GRBs. A natural problem will be raised as to whether short GRBs with plateau
phase in the afterglow also obey the same correlation. Unfortunately, the number of short
GRBs meeting the requirement is currently too few.

It is worth noting that many interesting physics could be involved in newly born
magnetars (Dall$'$Osso et al. 2009). The tops include the emission of gravitational waves,
the cooling process, the evolution of the magnetic axis, etc. Some of the physics may
affect the the energy injection process of the newly born magnetar delicately. We believe
that further studies on the new three-parameter correlation may give useful constraints on
the physics of newly born magnetars.

\begin{acknowledgements}
We thank the anonymous referee for many of the useful suggestions and comments.
We also would like to thank Z. G. Dai, S. Qi, and F. Y. Wang for helpful discussion.
This work was supported by the National Natural Science Foundation of China (Grant
No. 11033002), and the National Basic Research Program of China (973 Program, Grant
No. 2009CB824800).

\end{acknowledgements}


\newpage

\begin{table*}
\label{tab:sample}
\begin{center}
{
\begin{tabular}{|c|c|c|c|c|c|}
\hline
GRB & $z$ & ${\rm Log}[L_{\rm X}/(\rm erg/s)]$ & ${\rm Log}[T_{\rm a}/(\rm s)]$
& ${\rm Log}[E_{\gamma,\rm iso}/(\rm erg)]$  & Type \\
\hline
050315	&	1.95	&	47.05 	$\pm$	0.19 	&	3.92	$\pm$	0.17	&	52.85 	$\pm$	0.012 	&	Long	\\
050319	&	3.24	&	47.52 	$\pm$	0.18 	&	4.04	$\pm$	0.17	&	52.90 	$\pm$	0.057 	&	Long	\\
050401	&	2.9	&	48.45 	$\pm$	0.15 	&	3.28	$\pm$	0.14	&	52.50 	$\pm$	0.098 	&	Long	\\
050416A	&	0.65	&	46.29 	$\pm$	0.23 	&	2.97	$\pm$	0.21	&	51.02 	$\pm$	0.027 	&	Long	\\
050505	&	4.27	&	48.03 	$\pm$	0.34 	&	3.67	$\pm$	0.33	&	53.26 	$\pm$	0.019 	&	Long	\\
050724	&	0.26	&	44.53 	$\pm$	1.24 	&	4.92	$\pm$	1.22	&	50.17 	$\pm$	0.055 	&	IC	\\
050730	&	3.97	&	48.68 	$\pm$	0.07 	&	3.44	$\pm$	0.04	&	53.26 	$\pm$	0.017 	&	Long	\\
050801	&	1.38	&	47.86 	$\pm$	0.17 	&	2.17	$\pm$	0.16	&	51.49 	$\pm$	0.066 	&	Long	\\
050802	&	1.71	&	47.43 	$\pm$	0.06 	&	3.52	$\pm$	0.06	&	52.59 	$\pm$	0.021 	&	Long	\\
050803	&	0.42	&	46.55 	$\pm$	0.87 	&	2.74	$\pm$	0.81	&	51.46 	$\pm$	0.069 	&	Long	\\
050814	&	5.3	&	47.88 	$\pm$	0.47 	&	3.13	$\pm$	0.45	&	53.29 	$\pm$	0.029 	&	Long	\\
050824	&	0.83	&	45.30 	$\pm$	0.29 	&	4.65	$\pm$	0.27	&	51.13 	$\pm$	0.052 	&	Long	\\
050922C	&	2.2	&	48.92 	$\pm$	0.07 	&	2.08	$\pm$	0.07	&	52.77 	$\pm$	0.009 	&	Long	\\
051016B	&	0.94	&	47.59 	$\pm$	0.57 	&	3.22	$\pm$	0.55	&	51.01 	$\pm$	0.034 	&	Long	\\
051109A	&	2.35	&	48.01 	$\pm$	0.13 	&	3.4	$\pm$	0.11	&	52.72 	$\pm$	0.018 	&	Long	\\
051109B	&	0.08	&	43.51 	$\pm$	0.21 	&	3.64	$\pm$	0.19	&	48.55 	$\pm$	0.064 	&	Long	\\
051221A	&	0.55	&	44.74 	$\pm$	0.16 	&	4.51	$\pm$	0.16	&	51.40 	$\pm$	0.014 	&	IC	\\
060108	&	2.03	&	46.50 	$\pm$	0.13 	&	3.92	$\pm$	0.13	&	51.94 	$\pm$	0.027 	&	Long	\\
060115	&	3.53	&	47.80 	$\pm$	0.57 	&	3.09	$\pm$	0.55	&	52.99 	$\pm$	0.023 	&	Long	\\
060116	&	6.6	&	49.37 	$\pm$	0.33 	&	1.8	$\pm$	0.3	&	53.33 	$\pm$	0.082 	&	Long	\\
060202	&	0.78	&	45.64 	$\pm$	0.23 	&	4.74	$\pm$	0.23	&	52.00 	$\pm$	0.040 	&	Long	\\
060206	&	4.05	&	48.65 	$\pm$	0.10 	&	3.15	$\pm$	0.1	&	52.79 	$\pm$	0.013 	&	Long	\\
060502A	&	1.51	&	47.27 	$\pm$	0.19 	&	3.85	$\pm$	0.21	&	52.59 	$\pm$	0.012 	&	IC	\\
060510B	&	4.9	&	47.39 	$\pm$	0.49 	&	3.78	$\pm$	0.48	&	53.64 	$\pm$	0.011 	&	Long	\\
060522	&	5.11	&	48.51 	$\pm$	0.33 	&	2.07	$\pm$	0.31	&	53.05 	$\pm$	0.026 	&	Long	\\
060604	&	2.68	&	47.24 	$\pm$	0.19 	&	3.98	$\pm$	0.18	&	52.21 	$\pm$	0.069 	&	Long	\\
060605	&	3.8	&	47.76 	$\pm$	0.09 	&	3.48	$\pm$	0.08	&	52.66 	$\pm$	0.034 	&	Long	\\
060607A	&	3.08	&	45.68 	$\pm$	2.75 	&	4.14	$\pm$	0.02	&	53.12 	$\pm$	0.012 	&	Long	\\
060614	&	0.13	&	43.93 	$\pm$	0.05 	&	5.01	$\pm$	0.05	&	51.32 	$\pm$	0.006 	&	IC	\\
060707	&	3.43	&	48.01 	$\pm$	0.40 	&	2.94	$\pm$	0.36	&	52.93 	$\pm$	0.025 	&	Long	\\
060714	&	2.71	&	48.22 	$\pm$	0.08 	&	3.11	$\pm$	0.07	&	53.06 	$\pm$	0.016 	&	Long	\\
060729	&	0.54	&	46.17 	$\pm$	0.04 	&	4.73	$\pm$	0.04	&	51.69 	$\pm$	0.021 	&	Long	\\
060814	&	0.84	&	46.69 	$\pm$	0.06 	&	4.01	$\pm$	0.06	&	52.97 	$\pm$	0.004 	&	Long	\\
060906	&	3.69	&	47.73 	$\pm$	0.13 	&	3.62	$\pm$	0.12	&	53.26 	$\pm$	0.042 	&	Long	\\
060908	&	2.43	&	48.24 	$\pm$	0.11 	&	2.46	$\pm$	0.09	&	53.03 	$\pm$	0.010 	&	Long	\\
060912A	&	0.94	&	46.37 	$\pm$	0.23 	&	2.97	$\pm$	0.18	&	51.91 	$\pm$	0.020 	&	IC	\\
061121	&	1.31	&	48.35 	$\pm$	0.10 	&	3	$\pm$	0.09	&	53.47 	$\pm$	0.004 	&	Long	\\
070110	&	2.35	&	48.25 	$\pm$	0.72 	&	1.89	$\pm$	0.37	&	52.90 	$\pm$	0.033 	&	Long	\\
070208	&	1.17	&	46.88 	$\pm$	0.15 	&	3.63	$\pm$	0.14	&	51.58 	$\pm$	0.060 	&	Long	\\
070306	&	1.49	&	47.07 	$\pm$	0.05 	&	4.42	$\pm$	0.04	&	53.18 	$\pm$	0.008 	&	Long	\\
\hline
\end{tabular}
}
\end{center}
\caption{
55 GRBs of our sample. Data of the second, third, and forth columns are taken from D2010,
the fifth column, $E_{\gamma,\rm iso}$, are calculated from Equation (2),
where $S_{\rm bolo}$ are taken from Wang et al. (2011).
The last column is the type of the GRB, where Long means long GRB and IC is intermediate class GRB.
All the error bars are $1 \sigma$ range.
}
\end{table*}

\newpage
\begin{table}
\label{tab:sample}
\begin{center}
{
\begin{tabular}{|c|c|c|c|c|c|}
\hline
GRB & $z$ & ${\rm Log}[L_{\rm X}/(\rm erg/s)]$ & ${\rm Log}[T_{\rm a}/(\rm s)]$
& ${\rm Log}[E_{\gamma,\rm iso}/(\rm erg)]$  & Type \\
\hline
070506	&	2.31	&	47.63 	$\pm$	1.42 	&	2.87	$\pm$	1.42	&	51.82 	$\pm$	0.029 	&	Long	\\
070508	&	0.82	&	48.20 	$\pm$	0.02 	&	2.75	$\pm$	0.02	&	53.11 	$\pm$	0.004 	&	Long	\\
070529	&	2.5	&	48.40 	$\pm$	0.15 	&	2.34	$\pm$	0.15	&	53.04 	$\pm$	0.025 	&	Long	\\
070714B	&	0.92	&	46.85 	$\pm$	0.20 	&	3.03	$\pm$	0.19	&	52.30 	$\pm$	0.033 	&	IC	\\
070721B	&	3.63	&	47.08 	$\pm$	0.51 	&	3.58	$\pm$	0.51	&	53.34 	$\pm$	0.035 	&	Long	\\
070802	&	2.45	&	46.84 	$\pm$	2.72 	&	3.68	$\pm$	0.62	&	51.96 	$\pm$	0.047 	&	Long	\\
070809	&	0.22	&	44.15 	$\pm$	0.76 	&	4.09	$\pm$	0.75	&	49.43 	$\pm$	0.062 	&	IC	\\
070810A	&	2.17	&	47.97 	$\pm$	0.13 	&	2.83	$\pm$	0.12	&	52.26 	$\pm$	0.023 	&	IC	\\
071020	&	2.15	&	49.22 	$\pm$	0.05 	&	1.84	$\pm$	0.05	&	52.87 	$\pm$	0.016 	&	Long	\\
080310	&	2.42	&	46.72 	$\pm$	0.11 	&	4.08	$\pm$	0.11	&	52.88 	$\pm$	0.023 	&	Long	\\
080430	&	0.77	&	46.03 	$\pm$	0.08 	&	4.29	$\pm$	0.08	&	51.68 	$\pm$	0.022 	&	Long	\\
080603B	&	2.69	&	48.88 	$\pm$	0.26 	&	2.92	$\pm$	0.24	&	53.07 	$\pm$	0.011 	&	Long	\\
080810	&	3.35	&	48.24 	$\pm$	0.08 	&	3.28	$\pm$	0.07	&	53.42 	$\pm$	0.031 	&	Long	\\
081008	&	1.97	&	47.79 	$\pm$	0.24 	&	2.95	$\pm$	0.22	&	52.85 	$\pm$	0.047 	&	Long	\\
090423	&	8.26	&	48.48 	$\pm$	0.11 	&	2.95	$\pm$	0.1	&	53.03 	$\pm$	0.018 	&	Long	\\
\hline
\end{tabular}
}
\end{center}
\end{table}

\newpage

\end{document}